\documentclass{PoS}

\def\a{\alpha}
\def\b{\beta}

\def\l{\lambda}

\def\ue3{\left| U_{e3} \right|}

\def\be{\begin{equation}}
\def\ee{\end{equation}}

\PoS{PoS(HEP2005)187}

\title{Hybrid Textures of Neutrinos}

\ShortTitle{Hybrid Textures of Neutrinos}

\author{\speaker{Satoru Kaneko}
\\
Ochanomizu University, Japan\\
E-mail: \email{satoru@phys.ocha.ac.jp}}

\author{Hideyuki Sawanaka
\\
Niigata University, Japan\\
E-mail: \email{hide@muse.sc.niigata-u.ac.jp}}

\author{Morimitsu Tanimoto
\\
Niigata University, Japan\\
E-mail: \email{tanimoto@muse.sc.niigata-u.ac.jp}}

\abstract{
We present  numerical and comprehensive analyses of the sixty  hybrid textures
of neutrinos, which have an  equality of matrix elements and  one zero. 
These textures are possibly derived in the models with discrete flavor symmetry.
Only six textures among sixty ones  are excluded by the present experimental
data. Since there are many textures  which give similar predictions,
the textures are classified  based on  the numerical results.
The neutrinoless double beta decay is also examined in these textures.
Our results suggest  that there remain still rich structures of 
 the neutrino mass matrix in the phenomenological point of view.
 }

\FullConference{International Europhysics Conference on High Energy Physics\\
		 July 21st - 27th 2005\\
		 Lisboa, Portugal}
		 
\begin{document}


The results of neutrino oscillation experiments indicate the neutrino masses and mixings,    
especially, the bi-large flavor mixing. 
It is therefore important to investigate how the textures of lepton mass 
matrices can link up with the observables of the flavor mixings.
 Many authors studied the texture zeros
 \cite{Pre, FGM1, Xing}, which may follow from the flavor  symmetry.
 On the other hand,  one finds  some  relations
among the non-zero mass matrix elements 
in the discrete symmetry of the flavor.
This fact suggests that the texture zero analyses is not enough to reveal
  some underlying flavor symmetry.

For example, in the $e, \mu, \tau$ basis,
 one finds the following symmetric mass matrices of neutrinos, where
there are  two entries with same values and one zero,
\begin{equation}
\left (\matrix{ a &  c & d\cr  & b & 0 \cr &  & b\cr}
\right ) \ , \qquad 
\left (\matrix{ 0 &  c & d\cr  & a & b \cr  &  & a\cr}
\right ) \ ,
\end{equation}
\noindent
both are presented  in the quaternion family symmetry, $Q_8$ \cite{Q8}
 and the latter is given   in the $S_3$ symmetry \cite{S3}.
The variant of these textures is also discussed in ref. \cite{Koide}.
We call this type texture as the  ``Hybrid'' texture,
which has an   equality of matrix elements and  one zero. 

The analytical study of various structures of the neutrino mass matrix was
presented systematically by Frigerio and Smirnov \cite{FS}, who also discussed
the case of equalities of matrix elements.
The textures for the Dirac neutrinos also discussed in  \cite{Hage}.
However, numerical and comprehensive analyses have not been given.  
In this work, we present numerical  analyses of 
the  sixty hybrid textures,  which have 
   equal two  neutrino mass matrix elements with one zero 
   \cite{hybrid}.
Our analyses include textures in the previous studies in \cite{Q8,S3}.
Our results  are consistent with their ones.




Let us construct the neutrino mass matrix in terms of neutrino mass 
eigenvalues $m_1,m_2,m_3$, mixing angles  and CP violating phases.
Neutrino mass matrix  $M_\nu$ 
     in the basis where the charged lepton mass matrix  is diagonal 
 (flavor basis)  is  given as follows:
\begin{equation}
 M_\nu= P \ U^* \ M_{\rm diagonal}  \ U^\dagger \ P \ ,
\end{equation}
\noindent where  $M_{\rm diagonal}$ and $P$ are  the diagonal mass matrix and
  the diagonal phase matrix, respectively:
\begin{equation}
M_{\rm diagonal}=
{\rm diag}(\l_1,\l_2,\l_3)\ ,\ \ 
P = {\rm diag}(e^{i \phi_e},e^{i \phi_\mu},e^{i\phi_\tau})\ ,
\end{equation}
\noindent
where $\l_i\ (i=1,2,3)$ are the complex mass eigenvalues including 
Majorana phases, so three neutrino masses $m_i$ are given as 
the absolute values of $\lambda_i$. 
$U$ is the MNS mixing matrix. 
On the other hand, $\phi_i\ (i=e,\mu,\tau)$ are unphysical phases 
depending on the phase convention. 
Then, the neutrino mass matrix elements  ${(M _\nu)}_{\alpha\beta}$
are given in the flavor basis as
\begin{eqnarray}
 (M _\nu)_{\alpha\beta} =
 e^{i(\phi_\a+\phi_\b)}\sum _f^3 U _{\alpha f}^* U_{\beta f}^* \lambda _f = 
e^{i(\phi_\a+\phi_\b)}\ (U_{\alpha 1}^* U_{\beta 1}^* \lambda_1+U_{\alpha 2}^* U_{\beta 2}^*\lambda _2 + U_{\alpha 3}^* U_{\beta 3}^* \lambda _3)\ .
\end{eqnarray}
By use of these mass matrix elements, we can analyze the sixty  cases
in terms of mass eigenvalues and mixings.
These cases are combinations of the one zero textures  and 
the equal  elements:
 (i) fifteen  cases of the equal  elements (Type $\rm A\sim O$)
$(M _\nu)_{\a\b} = (M_\nu) _{\gamma\delta}$,
 (ii) six  cases of one zero (Type $\rm I\sim VI$)
$(M _\nu)_{\alpha\beta} = 0$.
These combinations give ninety textures, however, among them, thirty textures
have two zeros, which have been studied in details 
\cite{Pre,FGM1, Xing}.
Therefore, we study numerically in the sixty textures, 
which are summarized in \cite{hybrid}.


Since the textures have the conditions  $(M_\nu)_{\a\b}=0$ and
 $(M_\nu)_{ij}=(M_\nu)_{k\ell}$,
 we can get the ratios of mass eigenvalues by
solving two equations as follows \cite{Xing}:
\begin{eqnarray} 
\frac{\lambda_1}{\lambda_2} & = & 
\frac{(U_{i 3}^*U_{j 3}^*- Q\ U_{k3}^*U_{\ell 3}^*)  U_{\a 2}^* U_{\b 2}^* -
 (U_{i 2}^*U_{j2}^*- Q\ U_{k2}^*U_{\ell 2}^*)  U_{\a 3}^* U_{\b 3}^*}
{(U_{i 1}^* U_{j 1}^*- Q\ U_{k 1}^* U_{\ell 1}^*)  U_{\a 3}^* U_{\b 3}^* -
 (U_{i 3}^* U_{j3}^*- Q\ U_{k3}^* U_{\ell 3}^*)  U_{\a 1}^* U_{\b 1}^*}\ ,
\nonumber \\ \nonumber \\
\frac{\lambda_3}{\lambda_2} & = & 
\frac{(U_{i 1}^*U_{j 1}^*- Q\ U_{k1}^*U_{\ell 1}^*)  U_{\a 2}^* U_{\b 2}^* -
 (U_{i 2}^*U_{j2}^*- Q\ U_{k2}^*U_{\ell 2}^*)  U_{\a 1}^* U_{\b 1}^*}
{(U_{i 1}^* U_{j 1}^*- Q\ U_{k 1}^* U_{\ell 1}^*)  U_{\a 3}^* U_{\b 3}^* -
 (U_{i 3}^* U_{j3}^*- Q\ U_{k3}^* U_{\ell 3}^*)  U_{\a 1}^* U_{\b 1}^*}\ ,
\label{ratio}
\end{eqnarray}
 \noindent 
 where $ Q\equiv e^{i\varphi}= e^{i (\phi_k+\phi_\ell-\phi_i-\phi_j)}$.
Taking absolute values of these ratios, we get
the neutrino mass ratios, $m_1/m_2$ and $m_3/m_2$.
Therefore mass ratios are given in terms of  $\theta_{12}$,  $\theta_{23}$,
 $\theta_{13}$, CKM-like phase $\delta$ in $U$ and the unknown phase $\varphi$.
Absolute values of neutrino masses are fixed by putting the 
experimental data $\Delta m_{\rm atm}^2$ and $\Delta m_{\rm sun}^2$.


 We classify  the textures based  on the  predicted mixings
  $\theta_{23}$ and $|U_{e3}|$.
 We cannot  distinguish the textures  by the mixing $\theta_{12}$
 at the present stage of the experimental data: 
 \footnote{Please see \cite{hybrid} for the concrete textures of each type.}
\paragraph{Case 1 :}
The predicted mixings of the eighteen textures cover whole experimental 
allowed region : A-I, A-II, A-III, B-I, B-II, B-III, D-V, D-VI, E-II, F-I, I-I, I-II, 
   L-I, L-II, L-IV,  O-I, O-II and O-IV.

\paragraph{Case 2 :}
The allowed  points of the twenty-two textures  are very few : C-I, C-II, C-III, D-III,  E-V, E-VI, F-V, F-VI, G-III, G-V, G-VI, H-III, H-V, H-VI,  J-III, J-VI, K-III, K-VI, M-III, M-V,  N-III and N-V.

\paragraph{Case 3 :}
The  $\sin^2  2\theta _{23}$ has the lower bound $0.99$ : C-IV.

\paragraph{Case 4 :}
 The  $|U_{e3}|$ of the six textures has the lower bound, which increases 
 as $\sin^2 2\theta_{23}$ increases : A-VI, B-V,  G-II, H-I,  L-VI and O-V.
The lower bound $|U_{e3}|\geq 0.03$ is obtained in  A-VI and  B-V, and $|U_{e3}|\geq 0.04$  is  clearly predicted in G-II and H-I.
The bound $|U_{e3}|\geq 0.05$ is roughly obtained although the allowed points are few  in L-VI and  O-V.

\paragraph{Case 5 :}
 The  lower bound of $|U_{e3}|$ decreases as $\sin^2 2\theta_{23}$ increases in the seven textures : D-IV,  I-V, I-VI, J-II, K-I, M-II and N-I, in which $|U_{e3}|=0$ is allowed  at $\sin^2 2\theta_{23}=1$ except D-IV.  The texture  D-IV has the lower bound $|U_{e3}|\geq 0.001$.

\paragraph{Case 6 :}
The six textures are  excluded by the experimental data : E-IV,  F-IV,  J-IV, K-IV, M-IV and N-IV.



We will discuss a typical texture for each cases in the above classification :
\begin{eqnarray}
M_\nu
&=&
\left(
\begin{array}{ccc}
  X & 0  & e  \\
  &  X & f  \\
  &   &  c 
\end{array}
\right):\mbox{A-I}
\ ,\
\left(
\begin{array}{ccc}
  a & d  & e  \\
  &  X & 0  \\
  &   &  X
\end{array}
\right):\mbox{C-III}
\ ,\ 
\left(
\begin{array}{ccc}
  0 & d  & e  \\
  &  X & f  \\
  &   &  X 
\end{array}
\right):\mbox{C-IV}
\ ,\ 
\left(
\begin{array}{ccc}
  X & X  & 0  \\
  &  b & f  \\
  &   &  c 
\end{array}
\right):\mbox{G-II}
\ ,\ 
\nonumber\\
&&
\left(
\begin{array}{ccc}
  X & d  & e  \\
  &  0 & X  \\
  &   &  c 
\end{array}
\right):\mbox{I-V}
\ .\ 
\end{eqnarray}
The texture A-I is a typical one, which leads to the  normal hierarchy of 
the neutrino masses mainly, but the quasi-degenerate spectrum is also allowed.
The predicted mixings  cover all  experimental allowed
 region on  the  $\sin ^2 2\theta_{12}-|U_{e3}|$ plane
 as well as on  the  $\sin ^2 2\theta_{23}-|U_{e3}|$ plane.
The texture C-III  is a typical one, which leads to the inverted  
hierarchy of the neutrino masses mainly.
The texture C-IV  gives a specific mass hierarchy
 and  mixing angle $\theta_{23}$, 
 on the other hand,
  the predicted $\theta_{12}$ covers whole experimental allowed region.
The texture G-II is a typical one, which gives also 
a specific mass hierarchy
 and the  clear lower bound of  $|U_{e3}|$.
The texture I-V is a typical one, which leads to 
the inverted mass hierarchy of neutrino masses. 
The prediction excludes
the  specific region  on  the  $\sin ^2 2\theta_{23}-|U_{e3}|$ plane.

It may be helpful  to see which future data might rule out
these textures.
If the inverted mass hierarchy is shown to be realized by Nature,
 the textures of C-IV and G-II are ruled out.
On the contrary, if mass spectrum is the normal hierarchy,
the textures of  C-III and I-V are ruled out.
Finding $\sin^2 2\theta_{23}<0.98$ and $|U_{e3}|<0.04$ rule out
the texture C-IV and G-II, respectively.

It is important  to discuss the neutrinoless double beta decay rate, which 
 is controlled by the effective Majorana mass:
\begin{eqnarray}
 \langle m \rangle_{ee}=\left | \ m_1 c_{12}^2c_{13}^2 e^{i \rho} +
 m_2 s_{12}^2c_{13}^2 e^{i \sigma}+ m_3 s_{13}^2 e^{-2i \delta} \ \right | \ ,
\end{eqnarray}
\noindent where
$\rho=\arg{(\lambda_1/\lambda_3)}$ and $\sigma=\arg{(\lambda_2/\lambda_3)}$. 
This effective mass is just the absolute value of $(M_\nu)_{ee}$ component
 of the neutrino mass matrix. 
It is remarked that the neutrinoless double beta decay is forbidden 
in the textures of type IV, because of $(M_\nu)_{ee}=0$.
 Many hybrid textures (thirty-eight ones) predict 
the lower bound $10\sim 30$ meV although there are differences of factor 
in the lower bound predictions for each texture.

\end{document}